\documentclass[aps,pra,reprint, amsmath, amssymb,superscriptaddress,nofootinbib]{revtex4-1}

\usepackage{bm}
\usepackage[retainorgcmds]{IEEEtrantools}
\usepackage{graphicx}
\usepackage{mathrsfs}
\usepackage{amsmath}
\usepackage{amssymb}
\usepackage{color}
\usepackage{amsfonts}
\usepackage{times,txfonts}
\usepackage{nicefrac}
\usepackage[colorlinks=true,linkcolor=blue,urlcolor=blue,citecolor=blue,pdfusetitle]{hyperref}
\usepackage{amsmath}
\DeclareMathOperator{\sign}{sign}

\begin{document}

\title{Thermodynamic Skewness Relation From Detailed Fluctuation Theorem}
\date{\today}
\author{Domingos S. P. Salazar}
\affiliation{Unidade de Educa\c c\~ao a Dist\^ancia e Tecnologia,
Universidade Federal Rural de Pernambuco,
52171-900 Recife, Pernambuco, Brazil}

\begin{abstract}
The detailed fluctuation theorem (DFT) is a statement about the asymmetry in the statistics of the entropy production. Consequences of the DFT are the second law of thermodynamics and the thermodynamics uncertainty relation (TUR), which translate into lower bounds for the mean and variance of currents, respectively. However, far from equilibrium, mean and variance are not enough to characterize the underlying distribution of the entropy production. The fluctuations are not necessarily Gaussian (nor symmetric), which means its skewness could be nonzero. We prove
that the DFT imposes a negative tight lower bound for the skewness of the entropy production as a function of the mean. As application, we check the bound in the heat exchange problem between two thermal reservoirs mediated by a qubit swap engine.
\end{abstract}
\maketitle{}

%\section{Introduction}

% {\bf \emph{Introduction -}}
{\bf \emph{Introduction -}} 
Fluctuation Theorems (FTs) unlocked a variety of new results in nonequilibrium thermodynamics \cite{RevModPhys.93.035008,Campisi2011,Bustamante2005,Esposito2009,Jarzynskia2008,Jarzynski1997,Jarzynski2000,Crooks1998,Gallavotti1995,Evans1993,Hanggi2015,Saito2008,PhysRevX.11.031064}. In this context, the entropy production $\Sigma$ plays a major role as it is connected to different thermodynamic observables, such as irreversible work and heat transfer. In small systems far from equilibrium, the entropy production behaves randomly with fluctuations that satisfy some form of FT.

Among the FTs, the strong Detailed Fluctuation Theorem (DFT) \cite{Merhav2010,Seifert2012,Lupos2013,Andrieux2009} is a relation about the asymmetry of the probability density function of the entropy production between positive and negative values of $\Sigma$,  
\begin{equation}
\label{DFT}
   \frac{p(\Sigma)}{p(-\Sigma)}=e^{\Sigma}.
\end{equation}

The DFT arises in different situations, as observed in the setup of the exchange fluctuation theorem \cite{Jarzynski2004a,Timpanaro2019,Seifert2005,Garcia2010,Cleuren2006,Andrieux2009}, the Evan-Searles fluctuation theorem \cite{Evans2002,Crooks1999} and Gallavotti-Cohen relation \cite{Gallavotti1995}. The DFT (\ref{DFT}) shows that a positive entropy production is more likely to be observed when compared to the negative counterpart.

Some constraints for the statistics of $\Sigma$ are derived directly from the DFT. For instance, the most famous constraint that follows from (\ref{DFT}) is certainly the second law of thermodynamics \cite{RevModPhys.93.035008},
\begin{equation}
\label{secondlaw}
    \langle \Sigma \rangle \geq 0,
\end{equation}
now expressed as the ensemble average over $p(\Sigma)$.

Another constraint for the statistics of $\Sigma$ is the Thermodynamic Uncertainty Relation (TUR) \cite{Barato2015,Gingrich2016,MacIeszczak2018,Pietzonka2015,Pietzonka2017,Hasegawa2019,Timpanaro2019b}. In the setup of the strong DFT (\ref{DFT}),  when written in terms of the entropy production, the TUR \cite{Merhav2010,Hasegawa2019,Timpanaro2019} is a lower bound for the coefficient of variation, $\sigma/\langle \Sigma \rangle$, derived solely from the DFT (\ref{DFT}) and it reads
\begin{equation}
\label{TUR}
    \frac{\sigma}{\langle \Sigma \rangle} \geq 
    \sinh\big(\frac{g(\langle \Sigma \rangle\big)}{2})^{-1},
\end{equation}
for $\sigma:=\langle (\Sigma-\langle \Sigma \rangle)^2 \rangle^{1/2}$ and $g(x)$ is the inverse of $x\tanh(x/2)$ for $x\geq 0$. The TUR found applications in classic and quantum systems \cite{RevModPhys.93.035008}, where it is typically written in terms of currents.

%Although the DFT creates a dependency between $p(\Sigma)$ and $p(-\Sigma)$, it still leaves infinite possibilities for the distribution $p(\Sigma)$. For instance, the DFT allows symmetric distributions, $p(\Sigma)=p(\langle \Sigma \rangle - \Sigma)$, with zero skewness, $ \langle (\Sigma - \langle \Sigma \rangle)^3 \rangle = 0$, as the Gaussian case, but it also allows asymmetric distributions. 
Note that the objects of the second law (\ref{secondlaw}) and the TUR (\ref{TUR}) are the mean and variance of the distribution of the entropy production, $p(\Sigma)$. Although the first and second cumulants are sufficient to define a Gaussian distribution, there are several cases in which the distribution of the entropy production (and currents) might display non-Gaussian or asymmetric behavior \cite{Merhav2010,Timpanaro2019b,Salazar2019,Salazar2021,Salazar2020}, which makes the higher order cumulants important. In this case, the dependency between $p(\Sigma)$ and $p(-\Sigma)$ imposed by the DFT might impact higher order cumulants, such as the skewness.

In that context, skewness has been used to characterize asymmetry of distributions in several situations. For instance, skewness was used to characterize current fluctuations in a semiconductor quantum dot \cite{Gustavsson2006} and dusty plasma \cite{Belousov2016}, the first passage time in stochastic thermodynamics \cite{Barato2015,Wampler2001}, and transport in multi terminal junctions \cite{Ptaszynski2022}. Particularly, in connection to fluctuation theorems, skewness and higher order moments were analyzed in relation to transport coeficients \cite{Saito2008,Utsumi2009} and quantum work close to equilibrium \cite{Scandi2020}.

In this paper, in the same spirit of the second law (\ref{secondlaw}) and the TUR (\ref{TUR}), we ask if the strong DFT also influences the third central moment of the entropy production, $\langle (\Sigma -\langle \Sigma \rangle)^3\rangle$. In other words, assuming the strong DFT (\ref{DFT}) and given a known mean $\langle \Sigma \rangle$, can the distribution $p(\Sigma)$ be arbitrarily skewed? Or is there a thermodynamic skewness relation? In fact, we show that the skewness of the entropy production is bounded by the mean:
\begin{equation}
\label{TSR}
     \frac{\langle (\Sigma-\langle \Sigma \rangle)^3\rangle}{\sigma^3}\geq 
     -2\sinh\big(\frac{g(\langle \Sigma \rangle\big)}{2}).
\end{equation}

It means that the DFT (\ref{DFT}) allows the skewness of the distribution $p(\Sigma)$ to be positive, null (as in the Gaussian case) or negative, but not arbitrarily negative. The bound is tight and saturated by the same distribution that saturates both the TUR (\ref{TUR}) and the bound for apparent violations of the second law \cite{Salazar2021b}.

The paper is organized as follows. First, we present the formalism, where we define the skewness and prove that it is lower bounded by the mean (\ref{TSR}) using a technique based on Jensen's inequality. Alternatively, we also show that the skewness is not upper bounded by a function of the mean presenting an explicit counterexample. Then, as applications, we show how the entropy production of a nonequilibrium system operating in finite time behaves when compared to the lower bound: the entropy produced by a qubit swap engine. Finally, we discuss the results and some perspectives.

%\section{Formalism}
{\bf \emph{Formalism -}}
We prove a general result for the statistics of $p(\Sigma)$, assuming that $p(\Sigma)$ satisfies the strong DFT (\ref{DFT}). Similar strategies have been used in the TUR \cite{Merhav2010,Timpanaro2019b,Campisi2021,Y.Zhang2019} and apparent violations of the second law \cite{Salazar2021b}. First, let the skewness be defined as
\begin{equation}
\label{TSR0}
    \frac{\langle (\Sigma-\langle \Sigma \rangle)^3\rangle}{\sigma^3}=\frac{\langle \Sigma^3\rangle}{\sigma^3} -3\frac{\langle \Sigma \rangle}{\sigma} - \frac{\langle \Sigma \rangle^3}{\sigma^3}.
\end{equation}

The averages $\langle . \rangle$ are taken with respect to $p(\Sigma)$. Now we use a known property of odd functions under the DFT \cite{Salazar2021b,Hasegawa2019,Salazar2021}: let $u(\Sigma)$ be an odd function, $u(\Sigma)=-u(\Sigma)$, then we have the property
\begin{equation}
\label{oddproperty}
    \langle u(\Sigma) \rangle = \langle u(\Sigma) \tanh(\Sigma/2)\rangle.
\end{equation}
By setting the specific odd function $u(\Sigma):=\Sigma^3$, we have from (\ref{oddproperty}):
\begin{equation}
\label{thirdmomentum0}
    \langle \Sigma^3 \rangle = \langle \Sigma^3\tanh(\Sigma/2)\rangle = \langle g(h(\Sigma))^2 h(\Sigma) \rangle=\langle w(h(\Sigma)) \rangle,
\end{equation}
for $w(h):=g(h)^2 h$ and $h(\Sigma)=\Sigma \tanh(\Sigma/2)$, $g(h(\Sigma))=|\Sigma|$ for any $\Sigma$. Using $w''(h) \geq 0$ (see Appendix), we have from Jensen's inequality, 
\begin{equation}
\label{Jensens}
\langle w(h(\Sigma)) \rangle \geq w(\langle h(\Sigma) \rangle),    
\end{equation}
and replacing (\ref{Jensens}) in (\ref{thirdmomentum0}) results in
\begin{equation}
\label{thirdmomentum}
    \langle \Sigma^3 \rangle \geq w(\langle h(\Sigma) \rangle) = g(\langle h(\Sigma) \rangle)^2\langle h(\Sigma)\rangle = g(\langle \Sigma \rangle)^2 \langle \Sigma \rangle,
\end{equation}
where we used (\ref{oddproperty}) for $\tilde{u}(\Sigma)=\Sigma$, $\langle \Sigma \rangle=\langle \Sigma \tanh(\Sigma/2)\rangle = \langle h(\Sigma) \rangle$. Finally, replacing (\ref{thirdmomentum}) in (\ref{TSR0}), we obtain
\begin{equation}
\label{TSR2}
     \frac{\langle (\Sigma-\langle \Sigma \rangle)^3\rangle}{\sigma^3}\geq \frac{\langle \Sigma \rangle^3}{\sigma^3}\Big(\frac{g(\langle \Sigma \rangle)^2}{\langle \Sigma \rangle^2} -1\Big) -3\frac{\langle \Sigma \rangle}{\sigma}.
\end{equation}
From definitions of $h$ and $g$, we have 
\begin{eqnarray}
    \label{auxiliary1}
    h(g(\langle \Sigma \rangle))=g(\langle \Sigma \rangle) \tanh(g(\langle \Sigma \rangle)/2))=\langle \Sigma \rangle,
\end{eqnarray}
since $h(g(\langle \Sigma \rangle))=\langle \Sigma \rangle$, for $\langle \Sigma \rangle\geq 0$ (\ref{secondlaw}), which can be conveniently rewritten as
\begin{eqnarray}
    \label{auxiliary2}
    \frac{g(\langle \Sigma \rangle)^2}{\langle \Sigma \rangle^2}-1=\sinh(g(\langle \Sigma \rangle)/2)^{-2}.
\end{eqnarray}
Replacing (\ref{auxiliary2}) in (\ref{TSR2}) results in
\begin{equation}
\label{TSR3}
     \frac{\langle (\Sigma-\langle \Sigma \rangle)^3\rangle}{\sigma^3}\geq 
     \frac{\langle \Sigma \rangle}{\sigma}\Big(\frac{\langle \Sigma \rangle^2}{\sigma^2}\sinh(g(\langle \Sigma \rangle)/2)^{-2}-3\Big).
\end{equation}
Now we rewrite the rhs of (\ref{TSR3}) using the shorthand notation  $y:=\langle \Sigma \rangle/(\sigma\sinh(g(\langle \Sigma \rangle)/2))$:
\begin{equation}
\label{TSR4}
     \frac{\langle (\Sigma-\langle \Sigma \rangle)^3\rangle}{\sigma^3}\geq 
     \sinh(\frac{g(\langle \Sigma \rangle)}{2})y(y^2-3).
\end{equation}
Finally, as $y>0$ and $\sinh(g(\langle \Sigma \rangle)/2) \geq 0$, we use $y(y^2-3)=-2+(y-1)^2(y+2) \geq -2$ in (\ref{TSR4}) and obtain
\begin{equation}
\label{TSR5}
     \frac{\langle (\Sigma-\langle \Sigma \rangle)^3\rangle}{\sigma^3}\geq 
     -2\sinh(\frac{g(\langle \Sigma \rangle)}{2}).
\end{equation}

{\bf \emph{Remarks -}} One could check by inspection that the bound is actually the skewness of the minimal distribution, $p(\Sigma)=(e^{-a/2}\delta(\Sigma+a)+e^{a/2}\delta(\Sigma-a))/(2\cosh(a/2))$, with a given average, $\langle \Sigma \rangle=a\tanh(a/2)=h(a)$, which means $a=g(\langle \Sigma \rangle)$. Interestingly, this distribution also saturates the TUR (\ref{TUR}) \cite{Merhav2010,Timpanaro2019b}. 
For the skewness, we have for the minimal distribution $\langle \Sigma^2\rangle = a^2$, $\sigma=a/\cosh(a/2)$ and $\langle \Sigma^3 \rangle=a^3\tanh(a/2)=a^2\langle \Sigma \rangle$. In this particular case, the skewness (\ref{TSR0}) is given by
\begin{equation}
    \frac{\langle (\Sigma - \langle \Sigma \rangle)^3)}{\sigma^3}=-2\sinh(a/2)=-2\sinh(g(\langle \Sigma \rangle)/2),
\end{equation}
which saturates the lower bound (\ref{TSR}). Therefore, for a given mean, the minimal distribution minimizes the skewness and the variance simultaneously.

The behavior of the bound near equilibrium, $\langle \Sigma \rangle \approx 0$, is given by
\begin{equation}
     \frac{\langle (\Sigma-\langle \Sigma \rangle)^3\rangle}{\sigma^3}\geq 
     -2\sinh(\frac{g(\langle \Sigma \rangle)}{2}) \approx -\sqrt{2\langle \Sigma \rangle},
\end{equation}
since $h(x)\approx x^2/2$ for $x\approx 0^{+}$, we have $g(x)\approx \sqrt{2x}$ and $\sinh(x)\approx x$.

The bound is particularly useful in situations where the entropy production is simply written in terms of a current, $J$, one has $\Sigma = A J$, where $A$ is some affinity coefficient. In these cases, we have a Thermodynamic Skewness Relation (TSR) for the current,
\begin{equation}
\label{currentbound}
    \sign(A)\frac{\langle (J - \langle J \rangle)^3 \rangle}{\sigma_J^3} 
    \geq 
     -2\sinh\big(\frac{g(A\langle J \rangle\big)}{2}),
\end{equation}
where $\sigma_J:=\langle (J - \langle J \rangle)^2\rangle^{1/2}$. Note that (\ref{currentbound}) is a lower (upper) bound for the skewness of the current if $A>0$ ($A<0$).

Since we mentioned upper bounds, another important aspect of the skewness (\ref{TSR0}) of distributions $p(\Sigma)$ satisfying the DFT (\ref{DFT}) is that, although it is lower bounded, it is not upper bounded for a given mean. In other words, the DFT allows the skewness to be arbitrarily large. For that purpose, consider the ansatz for $\Sigma \in \mathbb{R}$
\begin{equation}
\label{upper0}
    p(\Sigma)=\frac{C(\lambda)}{1+(\Sigma/\lambda)^4},
\end{equation}
for $\Sigma \geq 0$ and $\lambda>0$,  and 
\begin{equation}
\label{upper1}
    p(\Sigma)=\frac{C(\lambda)e^{\Sigma}}{1+(\Sigma/\lambda)^4},
\end{equation}
for $\Sigma < 0$ and some $C(\lambda)$ that normalizes $p(\Sigma)$. One can easily check that (\ref{upper0}-\ref{upper1}) satisfy the DFT (\ref{DFT}). Moreover, $p(\Sigma)$ has a finite mean and variance. One could set $\lambda$ to fix a given mean,
\begin{equation}
    \langle \Sigma \rangle = C(\lambda)\lambda^2\int_0^\infty \frac{x(1-e^{-\lambda x})}{1+x^4} dx,
\end{equation}
by a change of variables $\Sigma = \lambda x$. However, the third momentum diverges for any $\lambda>0$,
\begin{equation}
\label{upper3}
    \langle \Sigma^3 \rangle = C(\lambda)\lambda^ 4\lim_{L\rightarrow \infty}\int_0^L \frac{x^3(1-e^{-\lambda x})}{1+x^4}dx=\infty,
\end{equation}
which makes the skewness (\ref{TSR0}) unbounded for a given mean under the DFT (\ref{DFT}).

\begin{figure}[htp]
\includegraphics[width=3.3 in]{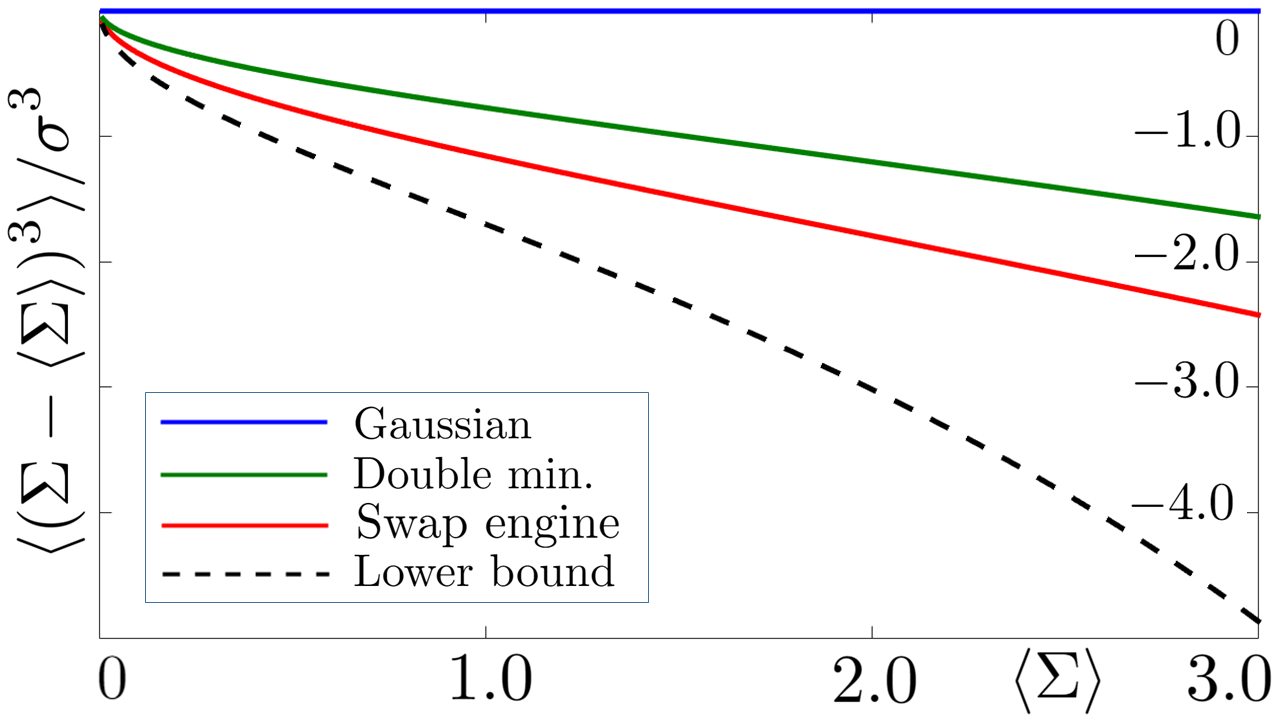}
\caption{(Color online) Examples of nonpositive skewness $\langle (\Sigma - \langle \Sigma \rangle)^3/\sigma^3$ as a function of the mean $\langle \Sigma \rangle$ for the swap engine (red), double minimal distributions (green) and Gaussian distribution (blue) and the lower bound (dashed), which is saturated by the minimal distribution. The curves collapse at $\langle \Sigma \rangle \approx 0$, where the bound behaves as $-\sqrt{2\langle \Sigma \rangle}$, which represents a near equilibrium situation.}
\label{fig1}
\end{figure}

{\bf \emph{Application to a swap engine -}} We consider a pair of qubits with energy gaps $\epsilon_A$ and $\epsilon_B$, initially prepared in thermal equilibrium, $p(\pm)=\exp(\pm\beta \epsilon)/(\exp(-\beta\epsilon)+\exp(+\beta\epsilon))$, for $\beta\in\{\beta_1,\beta_2\}$ and $\epsilon\in\{\epsilon_A,\epsilon_B\}$, with thermal reservoirs at temperature $T_1$ and $T_2$. The energy is measured in a two point measurement scheme performed before and after a swap operation \cite{Campisi2015}, where the swap operation is defined as $|xy\rangle \rightarrow |yx\rangle$, for $x,y \in \{-,+\}$. The entropy production is given by \cite{Campisi2015,Timpanaro2019b} by $\Sigma = \beta_1 \Delta E_A + \beta_2 \Delta E_B$,
where $\Delta E_A = E_A^f-E_A^i$, $\Delta E_B=E_B^f-E_B^i$ are the variations of energy measurements before and after the swap. The possible outcomes for the entropy production are $\Sigma \in s=\{0,\pm 2a\}$ for $2a=2(\beta_2\epsilon_B-\beta_1\epsilon_A)$ and the distribution $p(\Sigma)$ is given by $p(\Sigma)=(1/Z_0)\exp(\Sigma/2)$, for $\Sigma \in s$. It satisfies the DFT (\ref{DFT}), and it has simple relations for the statistical moments,
\begin{eqnarray}
\label{swap1}
    \langle \Sigma \rangle = \frac{a(e^{a/2}-e^{-a/2})}{1+e^{-a/2}+e^{a/2}},
    \\
    \label{swap2}
    \langle \Sigma^2 \rangle = \frac{a^2(e^{a/2}+e^{-a/2})}{1+e^{-a/2}+e^{a/2}},
    \\
    \label{swap3}
    \langle \Sigma^3 \rangle = \frac{a^3(e^{a/2}-e^{-a/2})}{1+e^{-a/2}+e^{a/2}}=a^2\langle \Sigma \rangle.
\end{eqnarray}
Finally, using expressions (\ref{swap1}-\ref{swap3}), we compute the mean (\ref{swap1}) and the skewness using (\ref{TSR0}) for different values of $a$, then we plot the skewness as a function of the mean $\langle \Sigma \rangle$ in Fig.~1.

{\bf \emph{Double minimal-}} We consider the double minimal distribution as another example, defined as the sum $\Sigma = \Sigma_1+\Sigma_2$, such that each $\Sigma_i$ satisfies the minimal distribution, $p(\Sigma_i)=(e^{-a/2}\delta(\Sigma_i+a)+e^{a/2}\delta(\Sigma_i-a))/(2\cosh(a/2))$, for $i=1,2$. Note that the possible outcomes for $\Sigma=\Sigma_1+\Sigma_2$ are $\{-2a,0,2a\}$ with corresponding probabilities $P(\Sigma=-2a)=p(\Sigma_1=-a)p(\Sigma_2=-a)=De^{-a}$, $P(\Sigma=2a)=p(\Sigma_1=a)p(\Sigma_2=a)=De^{a}$, $P(\Sigma=0)=p(\Sigma_1=-a)p(\Sigma_2=a)+p(\Sigma_1=+a)p(\Sigma_2=-a)=2D$, for $D=1/(e^{a}+e^{-a}+2)$. Check that it satisfies the DFT (\ref{DFT}). In this case, we obtain for the moments:
\begin{eqnarray}
\label{dmin1}
    \langle \Sigma \rangle = \frac{2a(e^{a}-e^{-a})}{2+e^{-a}+e^{a}},
    \\
    \label{dmin2}
    \langle \Sigma^2 \rangle = \frac{(2a)^2(e^{a}+e^{-a})}{2+e^{-a}+e^{a}},
    \\
    \label{dmin3}
    \langle \Sigma^3 \rangle = \frac{(2a)^3(e^{a}-e^{-a})}{2+e^{-a}+e^{a}}=(2a)^2\langle \Sigma \rangle.
\end{eqnarray}
Note that (\ref{dmin1}-\ref{dmin3}) resemble (\ref{swap1}-\ref{swap3}), with a slight difference that reflects in the skewness depicted in Fig.1. Also note that summing iid minimal random variables will make the skewness depart from the lower bound, as expected.

{\bf \emph{Gaussian case-}} For completeness and due to its importance, we include the Gaussian case,
\begin{equation}
\label{Gaussian}
    p(\Sigma)=\frac{1}{2\sqrt{\pi \langle \Sigma \rangle}}\exp\big(-\frac{(\Sigma - \langle \Sigma \rangle)^2}{4\langle \Sigma \rangle}\big),
\end{equation}
where the mean is given by $\langle \Sigma \rangle$, and it satisfies the DFT (\ref{DFT}), which actually fixes the variance in $\sigma^2=2\langle \Sigma \rangle$. Thus, differently from the general case, the Gaussian case has only one free parameter. Because it is symmetric around the mean, $p(\Sigma)=p(\langle \Sigma \rangle - \Sigma)$, then the skewness (\ref{TSR0}) of (\ref{Gaussian}) is zero for any $\langle \Sigma \rangle$, also depicted in Fig.1.

{\bf \emph{Discussion and Conclusions -}}
Using the strong Detailed Fluctuation Theorem (DFT), we showed that the skewness, here defined as $\langle (\Sigma -\langle \Sigma \rangle)^3\rangle/\sigma^3$, is lower bounded by the mean, $\langle \Sigma \rangle$. The lower bound is always negative, as depicted in Fig.~1. Near equilibrium, $\langle \Sigma \rangle$, the lower bound approaches $0$ as the function $-\sqrt{2\langle \Sigma \rangle}$.

The lower bound is saturated by a simple two level system, called the minimal distribution. It is minimal in the sense that it is the most simple distribution (two point mass function) satisfying the DFT with a given mean $\langle \Sigma \rangle$. We also showed an explicit example of a distribution satisfying the DFT (\ref{DFT}) with arbitrarily positive skewness (\ref{upper0}-\ref{upper1}). Thus, the DFT allows the the skewness to be arbitrarily positive, but not arbitrarily negative.

Interestingly, the same minimal distribution that saturates the skewness relation also saturates a form of Thermodynamic Uncertainty Relation (TUR) and the lower bound for apparent violations of the second law.

As applications, we considered the swap engine as a simple form of entropy production, with a distribution $p(\Sigma)$ given by a three point mass function. As expected, as it departs from the minimal distribution, Fig.~1 shows that the system has negative skewness, but it is above the lower bound as expected. Another example was the sum of two iid variables satisfying the minimal distribution, where the distribution of the sum is also a three point mass function. The Gaussian case (zero skewness) was also included for completeness.

As previous results derived from specific forms of Fluctuation Theorems, our result is limited by the applicability of the strong DFT. It might find straightforward applications, for instance, in the context of heat exchange (exchange fluctuation theorems). In this case, the bound written in terms of an observable current (\ref{currentbound}) and the underlying affinity is particularly useful.

{\bf \emph{Appendix-}} We prove that $w''(h)\geq0$, for $w(h):=g(h)^2 h$, where $h(x):=x\tanh(x/2)$ and $g(h(x))^2=x^2$, for any $x\in 	\mathbb{R}$.
We use the notation $w':=dw/dh$ and $\dot{w}:=dw/dx$.
From defition, we have $w=x^2h$ and $w'=\dot{w}x'$. The second derivative reads
\begin{eqnarray}
\label{app1}
    w''=\frac{d}{dh}(\dot{w}x')=\ddot{w}x'^2+\dot{w}x'',
\end{eqnarray}
where $x'=dx/dh=1/\dot{h}$ and $x''=x'(d/dx)(1/\dot{h})=-\ddot{h}x'/\dot{h}^2=-\ddot{h}/\dot{h}^3$. Replacing $x'$ and $x''$ in (\ref{app1}) yields 
\begin{eqnarray}
\label{app2}
    w''=\frac{1}{{\dot{h}^2}}\big(\ddot{w}-\dot{w}\frac{\ddot{h}}{\dot{h}}\big).
\end{eqnarray}
Finally, using $h(x)=x\tanh(x/2)$ and $w(h(x))=x^3\tanh(x/2)$ explicitly to calculate $\ddot{w}, \dot{w}$, $\dot{h}$ and $\ddot{h}$, one obtains
\begin{eqnarray}
\label{app3}
    w''=\frac{8\cosh(\frac{x}{2})^4(x^2+3(\cosh(x)-1)+3x\tanh(\frac{x}{2}))}{x^2(1+\sinh(x)/x)^3}.
\end{eqnarray}
We have from (\ref{app3}) that $w''>0$ for all $h\geq 0$. Actually, for $x\approx 0$, we have $w(h(x))\approx 4+x^2$.

\bibliography{lib4}
\end{document}